Working paper

# Demographic perspectives in research on global environmental change

Raya Muttarak (muttarak@iiasa.ac.at)

# Table of contents

ZVR 524808900

**Disclaimer:**

The author gratefully acknowledge funding from IIASA and the National Member Organizations that support the institute (The Austrian Academy of Sciences; The Brazilian Federal Agency for Support and Evaluation of Graduate Education (CAPES); The National Natural Science Foundation of China (NSFC); The Academy of Scientific Research and Technology (ASRT), Egypt; The Finnish Committee for IIASA; The Association for the Advancement of IIASA, Germany; The Technology Information, Forecasting and Assessment Council (TIFAC), India; The Indonesian National Committee for IIASA; The Iran National Science Foundation (INSF); The Israel Committee for IIASA; The Japan Committee for IIASA; The National Research Foundation of Korea (NRF); The Mexican National Committee for IIASA; The Research Council of Norway (RCN); The Russian Academy of Sciences (RAS); Ministry of Education, Science, Research and Sport, Slovakia; The National Research Foundation (NRF), South Africa; The Swedish Research Council for Environment, Agricultural Sciences and Spatial Planning (FORMAS); The Ukrainian Academy of Sciences; The Research Councils of the UK; The National Academy of Sciences (NAS), USA; The Vietnam Academy of Science and Technology (VAST).

# Abstract

Human population is at the centre of research on global environmental change. On the one hand, population dynamics influence the environment and the global climate system through consumption-based carbon emissions. On the other hand, health and wellbeing of the population is already being affected by climate change. The knowledge on population dynamics and population heterogeneity thus is fundamental in improving our understanding of how population size, composition and distribution influence global environmental change and how these changes affect subgroups of population differentially by demographic characteristics and spatial distribution. Existing theoretical concepts and methodological tools in demography can be readily applied to the study of population and global environmental change but the topic has remained less central in demographic research. However, the increasing relevance of demographic research on the topic coupled with availability and advancement in data and computing facilities have contributed to growing engagement of demographers in this field. In the past couple of decades, demographic research has enriched climate change research both in the analysis of the impact of population dynamics on the global climate system as well as the impact of climate change on human population. The key contribution is in moving beyond the narrow view that population matters only in terms of population size but putting a greater emphasis also on population composition and distribution through presenting both empirical evidence and advanced population forecasting accounting for demographic and spatial heterogeneity. Whilst the research on how population dynamics influence the environmental and climate system is relatively advanced in recent years, what is missing in the literature is the study that investigates how global environmental change affect current and future demographic processes and consequently population trends. If global environmental change does influence fertility, mortality and migration, the three key demographic components underlying population change, population estimates and forecast need to adjust from the climate feedback in population projections. Indisputably, this is the new area of research that directly requires expertise in population science and contribution from demographers.

## About the authors

**Raya Muttarak** is the Program Director of the IIASA Population and Just Societies (POPJUS) Program and Acting Research Group Leader of the Migration and Sustainable Development (MIG) research group. She is also the Director of Population, Environment and Sustainable Development at the Wittgenstein Centre for Demography and Global Human Capital a cooperation between IIASA, the University of Vienna, and the Austrian Academy of Sciences. (Contact: muttarak@iiasa.ac.at)

# Introduction

The past decade has witnessed a rise in climate change concern worldwide (Funk et al. 2020). Even in 2020 during when the COVID-19 pandemic has overshadowed all other emergencies, concerns about the threat of global climate change persist – as expressed by a median of 70% across the respondents surveyed in 14 high-income countries as compared to a median of 69% who reported concerns about the threat of the spread of infectious diseases (Fagan and Huang 2020). Similarly for environmental issues, the majority (71%) of the respondents surveyed in 2019-2020 in 20 middle- and high-income countries reported that they would prioritise environmental protection over job creation (Funk et al. 2020). In fact, the share of people who favour the protection of the environment has also risen since 2005/2006. Apart from the influence of recent major climate movements and strikes such as the Friday climate strike, increasing in frequency and intensity of extreme weather events including droughts, floods, hurricanes, heatwaves and forest fires all have contributed to stronger concern about environmental and climate change (Konisky et al. 2016; Zanocco et al. 2019).

Given the salience of climate and environmental issues and their urgency, it is natural for population science to embrace environment and climate change topics in their research agenda. Indeed, Hunter and Menken (2015, p. 24) argue that 'the time is ripe for population scientists to become more involved in research on climate change'. Demography as a discipline that studies population-related phenomena particularly change in population size, composition, distribution and characteristics in a systematic manner (Nam 1979) is highly relevant to environment and climate change issues. The entry point for demographers to research on the environment is conventionally related to population growth (Pebley 1998). This is dated back over 200 years ago following the Malthusian view that uncontrolled population growth which would eventually deplete natural resources can outstrip the earth's carrying capacity. With larger population size being seen as a major driver of environmental problems (Ehrlich and Ehrlich 1990), accordingly, earlier engagement of demographers in environmental-related issues was predominantly concentrated on population growth.

However, demographic processes are connected with the environment beyond population growth. As illustrated in Figure 1, human population are closely linked with the environmental system both through the impact of population dynamics on the environment and as an agent being affected by environmental changes. It is also possible that such changes in turn impact demographic processes through influencing fertility, mortality and migration patterns.

With respect to human impact on the environment, the number of population is positively associated with the demand for natural resources including fossil fuels, water and land since each person requires food and energy to survive. Not only does food production require substantial amount of water, but also energy. With more mouth to feed, agricultural revolution leads to changes in land use patterns from small-scale agriculture to large-scale commercial farming, which is energy-intensive. This in turn can affect pollution levels such as air pollution from burning of fossil fuels for both production and consumption. Carbon and greenhouse gas emissions as a result of fossil fuels burning subsequently contribute to the rise in global average temperature which has reached 1°C above preindustrial levels.

While population size is evidently positively associated with demand of natural resources, consumption patterns of individuals actually vary across the life cycle. Although residential energy use continues to rise with age, transportation energy use peaks around early 50s then declines at older ages (O'Neill and Chen 2002). Throughout the life course, it is estimated that for the US population, per capita emissions of $CO_2$ starts to decrease with age when a person reaches his or her late 60s (Zagheni 2011). With older population being less likely to be active and use less energy appliances and transportation, changing age structure due to population aging will consequently lead to a reduction in carbon emissions (Kluge et al. 2014; Liddle 2011, 2014; Liddle and Lung 2010; O'Neill, Liddle, et al. 2012). Apart from age structure, consumption also varies with other demographic characteristics such as gender and education. Women, for instance, typically have lower emissions than men because of lower meat consumption, less long distance travels and higher use of public transportation, among others (Räty and Carlsson-Kanyama 2010; Shaw et al. 2020). Likewise, educational attainment matters for carbon emissions because whilst increase in the level of education is positively associated with economic growth which implies higher emissions, concomitantly higher female educational attainment is associated with lower fertility and consequently slower population growth (O'Neill et al. 2020). Human impact on the environment and the global climate system therefore also depends on demographic structures and composition.

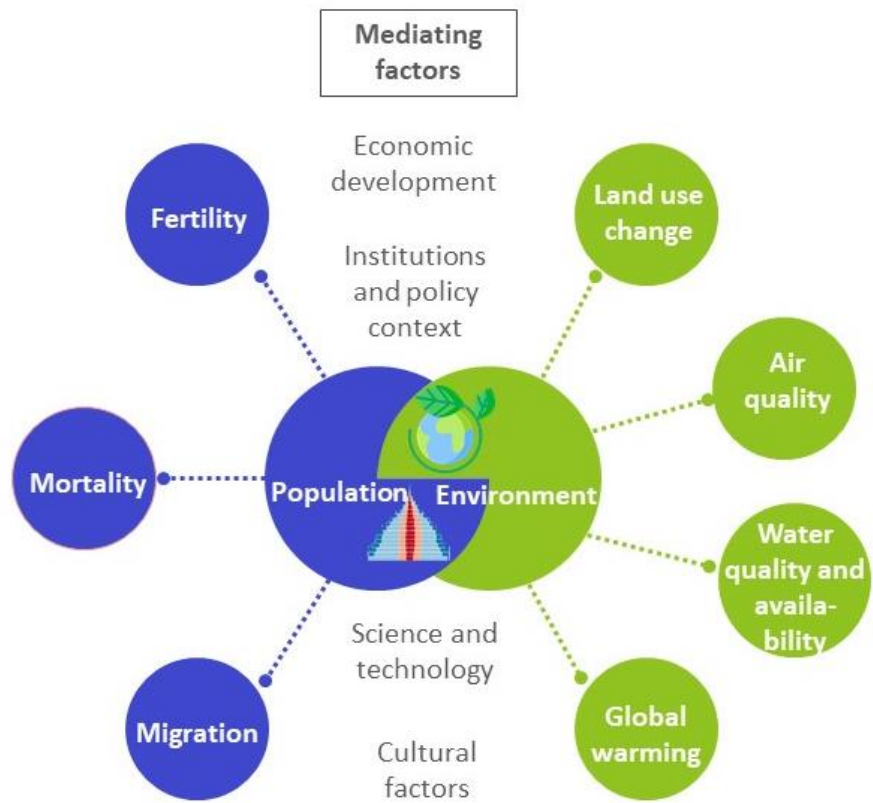

Source: Adapted from Hunter (2000, p. 4 (Figure 1.1)).

Figure 1: Conceptual framework describing the relationship between population and the environment

Similarly, population distribution is highly relevant to environmental and climate changes. It is estimated that urban areas are responsible for approximately 70% of greenhouse gas emissions (Johansson et al. 2012). Higher income of urban dwellers is associated with consumption-intensive lifestyles. Consumption of products and services (i.e. tangible and intangible goods such as education, healthcare, culture, recreation and restaurants) lead to indirect emissions beyond direct energy consumption (Ala-Mantila et al. 2014; Gill and Moeller 2018; Heinonen and Junnila 2011). However, due to higher population density, smaller dwellings and shorter transport distances, when considering per capita carbon footprint, cities do not necessarily emit more than the rural and suburban areas (Dodman 2009; Glaeser and Kahn 2010; Hoornweg et al. 2011). With the same level of wealth, it is found that urban areas in fact have the smallest footprints (Ala-Mantila et al. 2014). This suggests that the analysis of human impact on the environmental and climate systems need to account for both demographic and socioeconomic characteristics and population distribution.

Whilst population dynamics influences the environment and the global climate, at the same time human beings are affected by changes in the environmental and climate systems. The consequences of global warming on human health and wellbeing are already being felt. The past couple of decades have witnessed an increase in the frequency and intensity of extreme weather events in different areas of the world: for example, summer heatwaves in Europe, severe floods in India and Southeast Asia, severe droughts in the Sahel and Southern Africa, extreme rainfall from hurricanes in the US and forest fires in the US west coast and Australia, to name a few. These unprecedented weather shocks have been attributable to the increase in global average temperature induced by the accumulation of anthropogenic greenhouse gas emissions (Abatzoglou and Williams 2016; Harrington et al. 2019; Kew et al. 2019; Kumari et al. 2019; Oldenborgh et al. 2017; Otto et al. 2018). The changing climate affects the human population, for example, through changes in livelihoods, agricultural production, economic conditions, health and wellbeing (IPCC 2014a). Through these channels, it is likely that climate change will also affect demographic processes including fertility, mortality and migration and consequently future population size, distribution and composition. With the global temperatures on course of rising for a 2 to 5 degrees Celsius by the end of the century (Collins et al. 2013; Raftery et al. 2017), there is thus a potential feedback loop in human-environment systems whereby human activities impacts the environment and the changing environment in turn affects human population.

The impacts of environmental and climate change on human population however are not distributed evenly across population subgroups and the ability to adapt and cope with these changes varies substantially with population characteristics. Differences in physiological susceptibility, hazard exposure and socioeconomic and psychosocial factors influence risk perceptions and capacity to respond and this underlies demographically differentiated vulnerability (Muttarak et al. 2016). For instance, whilst mortality rates from tsunami are generally higher in women than men due to the lack of ability to swim (Neumayer and Plümper 2007) and the caregiving role prompting women to stay behind helping children and the elderly (Frankenberg et al. 2011; Yeh 2010), men are more likely to perish from floods and storms due to higher engagement in outdoor activities and higher risk-taking attitudes (Doocy, Daniels, et al. 2013; Doocy, Dick, et al. 2013; Zagheni et al. 2016). Since vulnerability to environmental change depends not only to the type of climatic hazards but also importantly on demographic characteristics, population composition hence is relevant to society's vulnerability and adaptive capacity (Lutz and Muttarak 2017).

Whilst demographic characteristics determine the degree of vulnerability and capacity to adapt to environmental change, climate risk also depends on population distribution. Exposure e.g. to extreme climate and weather events is one key component of risk (Cardona et al. 2012). If hazard events occur in an inhabited area, naturally no one is exposed to potentially harmful settings, and therefore there is no climate risk. Given higher density of population and high concentration of assets, urban centres are more susceptible to the risk and suffer greater casualties because of higher exposure. With approximately 55 percent of the world population living in the urban areas in 2018, demographic and geographic distribution of urban agglomerations coupled with socioeconomic and spatial vulnerabilities determines the risks posed by natural hazards (Gu 2019).

Vulnerability to climate change is also differentiated by the degree of susceptibility of each subpopulation. Whilst the urban areas are characterised by higher exposure to natural hazards, in terms of livelihoods people living in the rural areas depend heavily on climatic factors. Subsistence farmers relying on rainfall are particularly susceptible to climate change impacts and failure to adjust to climate variabilities can have serious consequences on their health and wellbeing. Child undernutrition, for instance, is found to increase with both droughts and floods, particularly in the rural areas and amongst agricultural households (Dimitrova 2020; Dimitrova and Muttarak 2020). In such households where crop yields are linked with food security, they are particularly susceptible to climate-induced extreme weather events and anomalies. Susceptibility is also closely linked with a physiological aspect. People of older age, for instance, are highly susceptible to extreme temperatures both during heatwaves and cold spells because of their low ability for thermoregulation (Baccini et al. 2008; Blatteis 2012; Kenny et al. 2017; Wanka et al. 2014). Meanwhile, infants and young children are susceptible to dehydration caused by diarrheal diseases which tend to increase after heaving rainfall and flooding events (Bennett and Friel 2014; Levy et al. 2016).

Hence, not only does *where* the population live matter, *who* is susceptible and to *what* hazard also underlies vulnerability. The impact of environmental and climate change on human populations thus depends not only population size but also substantially on distribution which determines exposure and composition of human populations which is linked with susceptibility, vulnerability and adaptive capacity.

The interactions between population and environment however are complex and are driven by many other forces. While population growth drives environmental change, consumption and production patterns are closely linked with socioeconomic development, which also determines population trends. The importance of considering the interconnectedness between systems including economic and social/cultural systems in understanding population-environment interactions has been highlighted by many scholars (Cohen 2010; Lutz 1994; Lutz, Fürnkranz-Prskawetz, et al. 2002; Martine 2005). The role of other factors in driving the impact of human activities on the environment and the level of vulnerability of human population to environmental changes is coined 'the sphere of the human-made environment' in Lutz et al. (2002, p. 4). People are seen as the agents who conduct their social and economic activities under different infrastructure, economy, government, politics, social structures, traditions, technology and information. Further development of this human-made environment thus influence the ultimate nature of the relationship between population and the environment.

Similarly, factors such as technologies, institutions and cultures are seen as 'mediating factors' in Hunter (2000, Chapter 5) where an extensive review of the role of these factors in mediating the relationship between human population dynamics and the natural environment is provided. No doubt, scientific and technological advancements influence production and consumption in all economic sectors ranging from agriculture and fishery, energy, waste management, construction, manufacturing to hospitality and tourism. Technological innovation such as irrigation technologies, development of new crop varieties, structural barriers for protection of coastal resources and floods and desalination has been introduced to facilitate adaptation to environmental change. Technological adoption in turn also depends on characteristics of the population which include culture. Culture influences the world view, values, beliefs and norms and consequently how human beings interact with the natural environment (Eisler et al. 2003; Liobikienė et al. 2016). Defined as the 'collective programming of the mind', culture 'distinguishes the members of one group or category of people from another' (Hofstede et al. 2010, p. 5). Cultural values such as son preference, gender role, individualistic culture and collectivist culture influence psychological processes underlying why and how individuals and groups engage in certain social behaviours which can have direct and indirect impact on the environment (e.g. through fertility behaviour, consumption level and preferences). For instance, whilst generally an increase in income is linked with a rise in meat consumption, this positive relationship is less steep in China and, particularly, India as compared to the United States (Ausubel and Gruebler 1995; Sans and Combris 2015). As a dietary habit is influenced by traditions and customs, almost one third of Indians follow lactovegetarianism (Devi et al. 2014). This shows that culture can alter the interactions between demographic and environmental factors.

Unquestionably, governance and institutions both at the local, national and international level are vital in mediating demographic pressure on the environment as well as mitigating the impact of environmental change. Whilst policy responses such as the Montreal Protocol of 1987 was successful in reducing the global emissions of chlorofluorocarbons (CFCs) through banning of products that contain CFCs (Hunter 2000), certain policy such as community relocation after natural disasters often exacerbate vulnerability of disadvantaged subgroups of population (Iuchi and Mutter 2020). Government sets up regulations and policies that monitor and incentivise uptake of mitigation and adaptation actions and thus contributes to mediating how people interact with the environment. The Shared Socioeconomic Pathways (SSPs) – descriptions of alternative futures of societal development, for instance, exemplify how different energy and socioeconomic development policies would yield varying trajectories of global warming as well as human vulnerability to environmental and climate change (O'Neill et al. 2014). Likewise, policy that may not be directly considered as environmental policy such as educational expansion may modify the human impact on the environment as well as contribute to vulnerability reduction (Lutz, Muttarak, et al. 2014; Lutz and Striessnig 2015).

The bidirectional relationship between population and the environment hence is determined by complex interactions among many other mediating factors and better scientific understanding of these interactions is called for (Hunter 2000).

In the remainder of the paper, the historical development of the field of population and environment is described with a focus on scrutinising why environmental and climate change research has not become central in population studies. The subsequent section explores the relevance and contribution of demography to climate change research both in terms of the impact of population dynamics on the climate system and the impact of climate change on the population. Through the extensive review of the development of the field, future directions on integrating demographic perspectives into global environmental change research is described. In particular, the research on the impact of climate change on population trends is highly relevant given the potential climate feedback on demographic processes themselves. The final section concludes.

# Historical development of the field of population and environment

With human population being central to the environmental and global climate systems, this makes demography a highly relevant field in environmental and climate research. Lutz et al. (2002) in a special issue in a *Population and Development Review* proposed that population-environment analysis deserves a specific field of study given a clearly identified unifying research question focusing on the impacts of population dynamics on the natural environment (P-E) on the one hand and the impacts of changes in the natural environment on human population (E-P) on the other.

While Lutz et al. (2002) argue that there has been an emerging critical mass of scholars working in this field, population and environment remains a minor field of study among demographers. This is evident in an international online survey of opinions and attitudes of 970 demographers who are a member of the International Union for the Scientific Study of Population (IUSSP) (Van Dalen and Henkens 2012). The study of environmental and climate issues was not listed amongst the key interests of demographers. Population aging was ranked as the top most important population issues facing the world in the next 20 years followed by mass migration, HIV/AIDS, above-replacement fertility, urbanisation and infant mortality. Despite the ever increasing prominence of climate change in the international community and publication debates (e.g. the joint award of Nobel Peace Prize to Intergovernmental Panel on Climate Change (IPCC) and former US Vice President Al Gore for their efforts to obtain and disseminate information about the climate challenge in 2007, the Paris Agreement adopted in 2015 and the Fridays for Future climate movement), research on environment and climate change remains peripheral in demographic research. This is reflected in very few sessions dedicated to population and environment in major demographic conferences such as annual meetings of the Population Association of America and European Population Conferences (Abel et al. 2018; McDonald 2016). The lack of engagement in environmental and climate-related research also results in the underrepresentation of demographers among major scientific efforts in climate change mitigation and adaptation actions including in the IPCC reports and the United Nations Climate Change Conferences.

# Why environmental and climate change issues have been peripheral in demographic research

The reasons why environmental and climate change issues remain less popular among demographers are summarised in Population Association of America Presidential Address published in *Demography* (Pebley 1998) and the special issue of the *Vienna Yearbook of Demography* which focuses on demographically differentiated vulnerability to climate-related disasters (Muttarak and Jiang 2015). Amongst the key reasons mentioned, the bitterness of the debate surrounding the issue 'limits to growth' plays a major role in influencing present day demographers in their engagement on environmental and climate change issues. Influenced by the Malthusian view that the continued growth of the population exponentially would exceed that of food production which grows arithmetically, population growth was viewed as a serious crisis starting from the late 1940s when the world population rose rapidly.

Subsequently, the late 1960s and early 1970s saw a rise in concerns about the capacity of the environment to absorb the multiple forms of pollution generated by population and economic growth. This is reflected in a series of publications including a book entitled *The Limits to Growth* published by the Club of Rome in 1972 (Ehrlich and Ehrlich 1970, Meadows et al. 1972). Analysing the interactions between the five basic factors underlying the earth's interlocking resources based on the data up to 1970, Meadows et al. (1972) show that as the world population grows, demand for material wealth increases leading to more resources use, higher industrial output and increasing pollution. The classic model used in this line of research proposed by Ehrlich and Holdren (1971) is the $I = P \times A \times T$ equation where $I$ stands for environmental impact, $P$ for population, $A$ for affluence and $T$ for technology. This equation has been operationalised and applied by considering $I$ as $CO_2$ emissions, $A$ as GDP per capita and $T$ as $CO_2$ per unit of GDP. Environmental impacts (e.g. emissions) thus were mainly considered as a function of the interactions between population size or growth, affluence and technology.

Continued population and economic growth therefore is not sustainable because the earth has a finite supply of resources. Accordingly, in the late 1960s and 1970s population control was seen as an essential mean to achieve economic growth and ensure sufficient environmental resources (McDonald 2016). Many scholars advocate fertility reduction as a mean to improve living standards and protect the environment and argued for population policy to be given immediate priority due to a time lag for the policy effects to realise (Jolly 1994). Accordingly, the focus of demographic research was on how to reduce fertility in less developed countries rather than on interactions between demographic factors and environmental variables (Pebley 1998). Given that the current debate on climate change is related to human activities and greenhouse gas emissions, the historical bitterness of this debate on population growth has made demographers be reluctant to engage with climate change and environmental issues (Gage 2016; Peng and Zhu 2016).

The second reason concerns the perception among many demographers that the research topics surrounding the environment and climate change such as production and consumption, technological advancement, regulations and institutions, disaster vulnerability and adaption are more directly related to other social science disciplines (e.g. economics, political science, geography) than demography (Hayes 2016; Pebley 1998; Peng and Zhu 2016). As a discipline focusing on a scientific study of human populations including the drivers of population dynamics and consequences of population change (IUSSP 2017), strictly speaking, the field of environment and climate change involves many other narratives which expand beyond formal demography (Peng and Zhu 2016). Keyfitz (1992) points out that the emphasis on the effects of other variables such as institutions, markets and feedback effects undermine the role of population dynamics in driving global environmental change. Therefore, instead of seeing population as closely linked to the environment and climate change, this topic was considered to be irrelevant for demographic research (Pebley 1998).

Furthermore, the complexity of climate and environmental science and the limitations of data and methods for integrating the environmental and climate context into the microdata commonly used by demographers are also barriers to more active engagement of demographers in this field (Hayes 2016; Hunter and Menken 2015). The study of global environmental change requires expertise in natural sciences as well as skills in spatial statistics, remote sensing and geographic information systems (Pebley 1998). Given that demography is a discipline with a strong empirical element (Caldwell 1996), the lack of appropriate data and analytical tools in the past hinder the involvement of demographers on this topic. The lack of engagement of demographers in the climate change research community results in the absence of social and demographic components in conventional climate models (e.g. the integrated assessment models (IAM) of the IPCC). Without consideration of societal and population change, it is not possible to provide integrated and reliable assessment of future change. Interdisciplinary collaboration can certainly fill this gap but differences in paradigms and assumptions in natural and social sciences remain an obstacle for cross-disciplinary fertilisation.

Finally, limitations in funding also make it challenging for demographers to take a new topic on board. Obtaining a research grant for a cross-cutting theme like population dynamics and environmental change has proven difficult, especially because a project proposal tends to still be evaluated within a traditional disciplinary specialisation. Despite a worldwide recognition of the urgency of interdisciplinary research to address global challenges and complex problems like climate change, there is evidence that interdisciplinary projects are less likely to be funded (Bromham et al. 2016). In addition, funders of climate change research tend to value natural sciences more than social science approaches (Peng and Zhu 2016). Overcoming this obstacle will facilitate demographers' involvement in this field.

Although the study related to the three components of population change, namely, fertility, mortality and migration remains a central focus of core demographic research, the past two decades have witnessed an increase in engagement of demographers in environmental and climate change research as well as widening interdisciplinary collaborations. The ratification of the Kyoto protocol in 1997 where industrialised countries were mandated to reduce their greenhouse gases emissions by 5.2% to a baseline of 1990 marked the first step in global efforts to tackle climate change (Böhringer 2003). Climate change mitigation has accordingly gained importance in international scientific and political debates. This marked the start of increasing engagement of demographers in research in the field of climate change.

One prominent contribution of demographers is in the development of the population-environment-development (PDE) approach – a system study of the complex interactions between population, development and environment under an integrated assessment framework (Lutz 1994; Lutz et al. 2004; Lutz, Fürnkranz-Prskawetz, et al. 2002; Lutz, Scherbov, et al. 2002; Lutz and Scherbov 2000; Sanderson 1994). Based on case studies, this approach combines a historical analysis of a case study country using qualitative and statistical methods with a series of simulation models. By producing different projections under alternative policy-relevant scenarios (e.g. population policy, policy on migration and remittances and energy policy), the PDE model is also relevant from a policy perspective allowing decision makers and stakeholders to explore alternative sustainable development paths through easy-to-use tools. In particular, the PDE employs the population-based approach which goes beyond a narrow view that only population growth or demographic changes matter for the environment (Lutz and Scherbov 2000). Human beings and their characteristics (such as age, sex, education, health and place of residence etc.) are considered as agents of changes in social, economic, cultural and environmental systems as well as agents whose well-being is directly affected by these changes. These demographic characteristics can be quantified and projected using the tools of multi-state demographic analysis and then integrated into systems analysis. The integrated dynamic systems help identify different policy options and consider how an intervention influences the interconnected human, social, economic and environmental systems in a scenario-based manner. The potential of demography in forecasting future population size, composition and distribution allowing for realistically matching future societies' characteristics with climate change scenarios (Lutz and Muttarak 2017).

Furthermore, the recent development in increasing availability of climate and natural disasters data in the public domain and repeated cross-sectional and longitudinal individual and household data containing demographic and relevant outcome variables opened up a new opportunity to study the interactions between population and environmental change (Fussell et al. 2014). Simultaneously, advancement in statistical techniques and computing technologies facilitate management and analysis of large-scale, complex environmental and demographic data. This new development in data and computational tools coupled with the urgency of environmental issues have encouraged participation of demographers in the field of environmental and climate change research.

# Relevance and contribution of demography in climate change research

That 'people are part of the problem of climate change and part of the solution' (Cohen 2010, p. 158) highlights the centrality of human population in the global climate system. As a discipline focusing on a scientific study of demographic trends and the drivers, naturally demography can provide insights into the demographic challenges in the context of climate change. Indeed, the Fourth Assessment Report of the Intergovernmental Panel on Climate Change (IPCC) explicitly called for contribution of social sciences, especially in improving the understanding of the social dimensions of climate change and vulnerabilities (IPCC 2014a). Furthermore, through more collaboration both within and across scientific fields, the World Social Science Report emphasizes the urgency of transforming social sciences into a "bolder", "better" and "bigger" field (ISSC and UNESCO 2013). In particular, the potential of social sciences to deliver solutions-oriented knowledge on the challenges posed by global environmental change is emphasized. This view has in fact already been put forward by Lutz (2012) who advocates for using social science study as "intervention sciences". In essential, intervention sciences refer to the social and economic sciences that study the current drivers of social change, how they will transform in the future and what actions (interventions) can alter the future pathway of events (Lutz 2012). Lutz and Striessnig (2015) argue that even though human behaviour is less deterministic than systems and phenomena studied in the natural sciences, it is still feasible to model social change where possible future outcomes are forecasted based on specified uncertainty range.

Understanding and predicting social change is of relevance in informing global environmental change policies and solutions which require models with predictive power for both the natural and social dimensions. However, scenario-based assessments commonly employ present-day socioeconomic conditions in assessing future biophysical impacts (IPCC 2014a). Despite the awareness that future societies will differ from those observed today, there remain little efforts to project alternative future scenarios of socioeconomic and human development (Lutz and Striessnig 2015). This is because it is rather difficult to quantify and develop alternative future scenarios in this field. Lutz and Muttarak (2017) nevertheless argue that certain aspects of societal development can be quantified and forecasted. Population dynamics, in particular, are highly relevant to both the anthropogenic impact on climate change and vulnerability and adaptation to climate change. Knowledge and methods in demography thus can be applied to improve our understanding of uncertainties, especially in the domain that is relevant to climate change mitigation, adaptive capacity and adaptation planning.

# Population impact on the climate

In the 1980s, it had become clear that global average temperature had increased and the planet is warming. Given a concern that this might be partly driven by human activities, the Intergovernmental Panel on Climate Change (IPCC) endorsed by the UN General Assembly was established in 1988. The mandate of the IPCC was to provide a comprehensive review and recommendations regarding research on climate change and its social and economic impact including identifying potential response strategies. Since 2001, following the publication of the IPCC Third Assessment Report, there has been a wide acceptance of anthropogenic climate change. That human activities cause changes to the climate system through burning of fossil fuels, land use change and consumption have highlighted the relevance of demography in understanding the human impact on the global climate system.

Traditionally, the role human population plays in carbon emissions was considered to be merely through population size or growth (de Sherbinin et al. 2007). When considering the three drivers of carbon emissions in the IPAT equation, moderating population growth (P) and low-carbon transformation (T) are viable policy options, given that intervention should not come at the expense of economic growth (A), especially for developing countries. Indeed, 12 percent of increased emissions in OECD countries between 1982 to 1997 was attributable to population growth alone (Hamilton and Turton 2002). Bongaarts et al. (1997) thus argued that by reducing global fertility only by half a birth per woman (corresponding with the United Nations' low variant population projection (United Nations 1992)), this only requires a 33 percent reduction in carbon intensity by the end of the century to keep total warming below 2°C. Given that slowing population growth through reducing fertility can contribute to emissions reduction, there has been a call for the climate community and particularly the IPCC to explicitly incorporate population policy into climate actions (Bongaarts et al. 1997; Bongaarts and O'Neill 2018). Although demographic factor is less important in determining short-term emissions as compared to per capita income and other variables, Lashof and Tirpak (1990) show that the population assumption is key in explaining the future path of greenhouse gas emissions by the year 2100. This makes the understanding of the world's changing demography highly relevant.

Population projections provide a tool to quantify uncertainty in future population trends underlying the future path of greenhouse gas emissions. The two most widely used long-range global population projections are provided by the United Nations (UN) Population Division made available since the early 1950s and by the World Population Programme of the International Institute for Applied Systems Analysis (IIASA) based outside Vienna (Austria) which began in 1994 (Lutz and KC 2010). A comprehensive review of different global population projections methods and assumptions can be found in O'Neill et al. (2001) and Lutz and KC (2010). The challenge in long-range population forecasting is how uncertainty is dealt with. This has important implication for climate change research. For instance, the two medium projections of the UN (United Nations 2015) and Wittgenstein/IIASA[1] (Lutz, Butz, et al. 2014) present a rather different outlook of the most likely future trend of world population with a difference of over two billion people. While the UN medium-variant projection from the probabilistic model forecasts that the world population will increase to 10.9 billion in 2100, the Wittgenstein/IIASA medium projection (most likely scenario) projects the increase of global population to 9.2 billion by the end of the century. Such large discrepancy between the two leading sources of demographic

---

[1] From 2011, IIASA global population projections are carried out in the framework of the Wittgenstein Centre for Demography and Global Human Capital (WIC) – a collaborative effort between the World Population Program of IIASA, the Vienna Institute of Demography of the Austrian Academy of Sciences (OEAW) and the Vienna University of Economics and Business (WU). In 2019, WU was no longer the university pillar of WIC and from 2020 is replaced by the Department of Demography, University of Vienna. Population projections

projections is not trivial when considering the impacts of human activities on the climate and environmental system.

Using a technique of probabilistic projection, the UN projection applies a Bayesian hierarchical model to estimate the double logistic curves for the total fertility rate and life expectancy at birth including probabilistic prediction intervals that give quantitative information about the range of uncertainty of future trajectories (Raftery et al. 2014). Probabilistic models present the likelihood of a future population value in the form of a probability distribution and thus yield only one output instead of multiple scenarios. Using the approach of expert-argument based projections, Wittgenstein/IIASA projections, on the other hand, rely on the online survey of 550 international experts including workshops focusing on future demographic trajectories of specific countries and world regions (Lutz, Butz, et al. 2014). The scientific inputs from the survey and workshops were synthesised and quantified providing numerical assumptions for the calculation of alternative global demographic scenarios to 2060 with extensions to 2100.

Whilst probabilistic projections within a short time horizon are reasonably accurate, there is no consensus on the reliable methods to generate reliable probabilistic bounds to represent the uncertainty of long-range population projections (Lutz and KC 2010; Rozell 2017). Wittgenstein/IIASA projections express uncertainty by using different scenarios which are linked to the Shared Socioeconomic Pathways (SSPs) developed by O'Neill et al. (2014). Providing the alternative pathways of global social and economic development over the next century, the SSPs have been used in the latest climate models incorporated in the IPCC sixth assessment report. Whilst the scenario-based approach has been criticised for the lack of quantification of uncertainty due to the absence of probability intervals and thus is not useful in providing a range of probable population size (Keilman 2020), Rozell (2017) argues that the Wittgenstein/IIASA projections are more suitable for analysing climate change policy options. By aligning the demographic scenarios with the SSPs scenarios for climate modelling used in IPCC reports, Wittgenstein/IIASA projections allow users such as policy makers to answer a type of 'What if…?' question in order to assess the effect of certain policy.

Apart from population size, other demographic processes and changes including population composition and distribution, ageing and urbanisation all have implications for consumption and production activities which in turn influence emissions driving climate change. A group of demographers have argued that research on drivers of climate change which focuses only on specific demographic variables (e.g. total population size) can misrepresent the effects of demographic change on emissions (Cohen 2010; de Sherbinin et al. 2007; O'Neill, Liddle, et al. 2012). Indeed, the knowledge on the spatial distribution of population by geographic regions and size of settlements is also fundamental in understanding land-use/land-cover change and greenhouse gas emissions. Given the established rural-urban differences in energy consumption, substantial urbanisation is found to be positively associated with per capita emissions (Liddle and Lung 2010; Parikh and Shukla 1995; York et al. 2003). In particular, higher income levels in larger metropolitan areas translate into higher consumption levels and emissions (Pachauri and Jiang 2008). Nevertheless, the impact of urbanisation on energy use and carbon emissions is not homogeneous for all countries. Exploring energy and emissions in India and China by a range of urbanisation scenarios, O'Neill et al. (2012), for example, show that because differences in per capita income between rural and urban areas are smaller in India in a baseline data, this explains smaller effects of urbanisation on emissions in India as compared to China.

Apart from the different levels of affluence and development, urban density is also a key factor underlying greenhouse gas emissions (Chen et al. 2008). The high concentrations of people and economic activities is found to be associated with lower levels of energy consumption and lower emissions in transport and buildings (Liddle 2014). Areas with high population density improve the efficient use of public infrastructure resulting in efficient public transport system and availability of local facilities and services. Despite the higher consumption due to higher wealth, since urban centres tend to have higher population density, it remains unclear whether urbanisation has overall positive or negative effects on greenhouse gas emissions. A recent study by Ribeiro et al. (2019) introduces a new approach which account for the confounding effects of the role of population size and population density including their interactions on urban emissions. It is found that apart from emissions in urban areas being dependent on population size and population density, larger cities in terms of population size and area also enjoy the larger impact of increasing population density on urban $CO_2$ emissions reduction. Demographic methods which account for spatial patterns of population changes thus are crucial in the estimation of greenhouse gas emissions.

Changing population composition is also highly relevant in forecasting future carbon emissions. O'Neill, Liddle et al. (2012) show that net of the effect of changes in population size, emissions in particular regions depend considerably on changes in population composition. In particular, given lower labour productivity of older populations which is consequently translated into a decline in economic growth, population aging is projected to contribute to emissions reduction in the long term by up to 20%. Apart from changing age structure, a recent study has also shown that changes in the education composition have implications on the climate system (O'Neill et al. 2020). The role of education on carbon emissions however is rather complex (Lutz and Striessnig 2015). On the one hand, increasing level of education leads to a desire for smaller family sizes which in turn contributes to slowing down population growth, especially in high-fertility contexts, and consequently lower emissions. On the other hand, education is associated with higher labour productivity and economic growth and as a consequence higher emissions. O'Neill et al. (2020) estimate for the first time the net education effects on emissions accounting for both its influence on population growth and economic growth. By the end of the century, a projected higher educational attainment under the SSP2 (Middle of the Road) scenario in all developing country regions lead to the net higher emissions due to the stronger role of education in promoting economic growth than its role in decreasing population. The readily available global population projections by age, sex and education under different SSP scenarios hence are highly useful in forecasting the impact of human activities on emissions (S. KC and Lutz 2017).

Future forecasts of energy consumption also depends on which units of analysis: individuals or households are used (MacKellar et al. 1995). Since a large portion of energy and energy-related commodities in both residential and transportation sectors are purchased and consumed jointly by household members, estimates and projections of household numbers are argued to be more relevant for forecasting energy consumption (de Sherbinin et al. 2007). Indeed, few cross-national studies show that emissions per person is negatively associated with average household size (Cole and Neumayer 2004; Liddle 2004). With lower energy consumption per person in larger households, a decline in household size following urbanisation and industrialisation across the globe can thus yield negative environmental impacts (Bradbury et al. 2014). Whilst there has been some progress in curbing population growth, growing number of households with a smaller household size highlights the importance of taking household dynamics into a study of demographic drivers of climate change.

By providing scientific insights into how current and future population size, distribution and composition drive climate changing carbon emissions (Gaffin and O'Neill 1997; Jiang and Hardee 2011; MacKellar et al. 1995; O'Neill et al. 2005, 2010; O'Neill, Liddle, et al. 2012), demography has made a significant contribution in the field of climate change.

# Impact of climate change on population

More recently, the interest in population dynamics in climate change research has also extended to identification of vulnerable populations and their locations through estimating the distribution and size of population potentially at risks of exposure to climatic hazards (de Sherbinin 2014; Harrington and Otto 2018; López-Carr et al. 2014). Population dynamics is unquestionably relevant to the understanding of hazard exposure and vulnerability. The Intergovernmental Panel on Climate Change (IPCC) in its recent Fifth Assessment Report defines vulnerability as 'The propensity or predisposition to be adversely affected. Vulnerability encompasses a variety of concepts and elements including sensitivity or susceptibility to harm and lack of capacity to cope and adapt.' (IPCC 2014b, p. 5). Population size is linked with vulnerability since rapid population growth and high-density population increase the number of people exposed to climate impacts and put pressure on the provision of basic services and infrastructure (Jiang and Hardee 2011).

The impacts of climate change however are not distributed evenly across population subgroups and the ability to adapt and cope with climate change varies substantially with population characteristics (Muttarak et al. 2016). Here demographic concepts and methods can be applied to identify demographically differentiated vulnerability which results from differences in physiological susceptibility, hazard exposure and socioeconomic and psychosocial factors influencing risk perceptions and capacity to respond. For instance, mortality rates from tsunami are generally higher in women than men due to the lack of ability to swim – physiological aspect underlying vulnerability – (Neumayer and Plümper 2007) and the caregiving role prompting women to stay behind helping children and the elderly – psychosocial factor underlying vulnerability (Frankenberg et al. 2011; Yeh 2010). On the other hand, men are more likely to perish from floods and storms due to higher engagement in outdoor activities and higher risk-taking attitudes – exposure and psychosocial factor underlying vulnerability (Doocy, Daniels, et al. 2013; Doocy, Dick, et al. 2013; Zagheni et al. 2016).

Apart from differentials in susceptibility and exposure, the ability to cope with climatic shocks also varies with demographic and socioeconomic characteristics. Vulnerability to climate hazards, for a woman, for instance, is not only associated with the differentials in physiological susceptibility but also with her socioeconomic position. A limited access to formal credit markets, for instance, makes female-headed households in South Africa more vulnerable to income loss due to climatic shocks (Flatø et al. 2017). A demographic characteristic may also interact with socioeconomic characteristics thus producing further differentials in vulnerability and adaptive capacity. Whilst generally education is reported to have a protective effect against vulnerability to climatic shocks (Muttarak and Lutz 2014), it is reported that mother's education has a stronger effect to mitigate drought-induced childhood undernutrition than father's education (Dimitrova 2020). Differential gender roles whereby women are more likely to be involved in childrearing may explain this finding. Children are reported to benefit from better health knowledge and access to healthcare of highly educated mothers that prevent them against being undernourished when the household experiences climatic shocks (Dimitrova and Muttarak 2020). Accounting for demographic differentials in coping capacity thus is fundamental for vulnerability reduction efforts.

In the past decade, following the inauguration of the Wittgenstein Centre for Global Demography and Human Capital in Vienna in 2010, a group of demographers have put forward education as an important source of population heterogeneity (S. KC et al. 2010; Lutz 2010; Lutz and KC 2011; Lutz and Skirbekk 2014). At the societal level, the role of education extends beyond influencing population dynamics. Changes in educational composition in a population are found to be key drivers of economic growth (Lutz et al. 2008, 2019), increase in life expectancy (Lutz and Kebede 2018) and even promoting democracy (Lutz et al. 2010). It has also been shown that providing quality education for all can promote achievement of the sustainable development agenda (S. E. L. Bengtsson et al. 2018; Lutz 2017). This argument is based on a series of evidence showing the (sometimes causal) links between education and other life domains such as labour market, health and gender equality. Given the importance of education in various desirable outcomes, unsurprisingly there is consistent evidence showing that education can contribute to reducing vulnerability and enhancing adaptive capacity (Muttarak and Lutz 2014).

The mechanisms through which education directly and indirectly influence vulnerability and adaptive capacity have been thoroughly discussed elsewhere (Dimitrova and Muttarak 2020; Hoffmann and Muttarak 2017; Muttarak and Lutz 2014). Muttarak (forthcoming) and Bengtsson et al. (2018) provide a comprehensive review of evidence on the role of education in reducing vulnerability in various domains. This ranges from education equipping individuals with better risk perceptions, higher disaster preparedness, lower morbidity and mortality, faster recovery from natural disasters to having more diversified adaptation options. Changing educational composition of the population can therefore influence the future impact of climate risks on population health and wellbeing (Lutz, Muttarak, et al. 2014; O'Neill et al. 2020).

With existing methodological tools to forecast future population distribution and composition, recent studies have combined empirical analysis of demographically differentiated vulnerability with multi-dimensional cohort component population projections to forecast future societies' vulnerability and adaptive capacity. For example, based on a regression analysis estimating mortality from climate-related disasters as an indicator of vulnerability (Striessnig et al. 2013), Lutz, Muttarak et al. (2014) project future disaster deaths by applying the education coefficients to the demographic scenarios underlying the Shared Socioeconomic Pathways (SSPs). SSPs scenarios are useful in this context given that different scenarios underlie different socioeconomic development pathways and their corresponding mitigation and adaptation challenges. Exploiting the readily available estimates of population size and composition (e.g. by age, sex and education) by SSPs scenarios until the year 2100 (S. KC and Lutz 2014, 2017), it is possible to perform the forecasting vulnerability of future societies exercise (Lutz and Muttarak 2017).

Climate risks do not only depend on the degree of vulnerability but also exposure. Rather than considering exposure as a static, simple description of geographical location associated with the risk of climate events, Martine and Schensul (2013) argue that exposure itself is shaped by a range of social and demographic processes. Apart from inquiring into '*what* place?' alone, it is necessary to consider '*who* is exposed?' and '*why* them?' (Martine and Schensul 2013, p. 11). Therefore, population characteristics need to be accounted for when analysing spatial population distribution. In order to provide meaningful projections of the spatial distribution of population that allows for identification of subpopulations vulnerable to the impact of climate change, demographic models that capture population heterogeneity at a smaller spatial resolution than a global scale are required.

Various spatial downscaling procedures with varying levels of complexity have been performed, especially in order to match with the scenario-based assessment of global change, future impacts, vulnerability and sustainable development (Zoraghein and O'Neill 2020). In particular, spatial projections that are consistent with the global change narratives describing future pathways of societal development such as the SSPs are highly relevant for policy planning. Unlike most existing spatially explicit global projections which employ simple scaling techniques or trend extrapolation (M. Bengtsson et al. 2006; Hachadoorian et al. 2011; van Vuuren et al. 2007), using the gravity-based population downscaling model developed by Jones and O'Neill (2013), Jones and O'Neill (2016) produce a set of global spatial population projections at a resolution of 1/8° (7.5 arc-minutes). National-level projections of urban and rural population change are downscaled in consistent with the demographic assumptions in each SSP narrative. A subsequent work by Gao (2017) further downscale the projections of Jones and O'Neill (2016) at a resolution of 1-km (about 30 arc-seconds) to match the need of some studies requiring data with a finer spatial resolution. These spatial projections of population can be matched with high resolution downscaled climate and hazard projections to identify future climate risk (B. KC et al. 2020; Smith et al. 2019).

Although spatial heterogeneity is well-captured in the spatially-explicit population projections, other dimensions of demographic heterogeneity have not been explicitly considered. In their recent work, Zoraghein and O'Neill (2020) produce a spatial population projections by rural and urban residence for each state in the United States at high resolution (1km). While this new approach has advanced our knowledge on rural and urban population change patterns of each US state, other demographic characteristics underlying vulnerability and adaptive capacity are not factored in. In an attempt to capture observable demographic dimensions of population heterogeneity, KC et al. (2018) develop a five-dimensional model of India's population by state, rural/urban place of residence, age, sex and level of education. It has been shown that the model that does not factor in education differentials and stratifying only by place of residence and state project the population size of India reaching 3.1 billion in 2100 as compared to 1.6 billion in the model stratifying only by level of education. The latter realistically accounts for the improvement in female education in India which is a key determinant of fertility rates. As previously discussed that education plays a key role in vulnerability reduction and enhancing adaptive capacity, the projections that do not consider such relevant demographic characteristics as education would undermine the efforts to identify vulnerable populations. This approach which focuses on population heterogeneity at the subnational scale allows for answering the questions of *who* is vulnerable and identifying *why* this is the case. However, refined spatial scales representing the local climate impact are not captured here.

Producing a meaningful set of population projections that simultaneously account for both spatial and demographic heterogeneity is highly challenging. Wardrop et al. (2018) propose a bottom-up approach for producing population estimates for small areas or high spatial-resolution grids. Relying on microcensus surveys collected for small, defined areas, the microcensus data are then linked to spatial covariate data using statistical models to predict population numbers also for unsampled locations. Exploiting the 2011 Population and Household Census and micro survey data (i.e. Demographic and Health survey (DHS) 2006 and 2011) for Nepal, KC et al. (2016) perform population projections for Nepal and its 75 districts for the period 2011-2031 by age and sex based on the cohort component method treating each district as a state in a multi-state framework. The results are further interpolated annually for 4,051 Village Development Committees (VDC)/Municipalities of Nepal. The technical possibility of producing multi-dimensional population projections at a small-scale spatial unit is promising. With increasing availability of subnational data on population and geo-referenced surveys coupled with computational power and advanced statistical technique, spatially disaggregated population estimation and projections which explicitly account for population heterogeneity would become a significant area of research (Wardrop et al. 2018). Expertise in demography is thus highly relevant here.

# Future directions in integrating demographic perspectives in global environmental change research

The impacts of climate change on human health and wellbeing have already been experienced by the population around the world. With the global temperatures on course of rising for a 2 to 5 degrees Celsius by the end of the century (Collins et al. 2013; Raftery et al. 2017), scaling up both mitigation and adaptation actions in order to reduce the human impact on the climate system and mimimise the climate risks should be made a priority. Given the centrality of human population in the global climate system and the urgency of climate change issues, there has never been 'the next best time for demographers to contribute to climate change research' (Gage 2016, p. 19). As the Chinese proverb says that the best time to plant a tree was 20 years ago, the time is now ripe for active engagement of demographers in environmental and climate study.

In particular, the past decade has seen significant advancement in demographic and social data and computational tools making complex environmental and climate data more accessible to demographers (Hunter and Menken 2015). The increasing availability of geo-referenced demographic data together with alternative data sources such as mobile phone and social media data make it possible to match them with climate and environmental data with appropriate spatial units and scales. This does not only allow researchers to provide better analysis and assessment of human impact on the global climate system and differential vulnerability and adaptive capacity but also ask new relevant questions in this field.

Back to Figure 1, which depicts the reciprocal relationship between population and the environment, the dynamics of climate changes are influenced by human population as much as population dynamics are influenced by the changing climate. It is reasonable to assume that climate change can also influence demographic processes given that the impact of climate change on human health and wellbeing is already being felt. The IPAT equation which describes the impact of human activity on the environment can be extended to consider the climate feedback on population trends. It can be written as:

$$I = P(I) \times A \times T$$

where *P(I)* refers to population which is a function of environmental and climate impact. Through impacting fertility, mortality and migration, the effects of climate change on the key demographic components successively influence future population size, distribution and composition which in turn have implications on demand for resources and the environment alike.

The notion that environmental conditions can influence demographic behaviour is not new. In their commentary piece arguing for the timely involvement of population scientists in climate change research, Hunter and Menken (2015) raise a relevant point on the already existing environmental aspects in classic demographic theories. Caldwell's classic theory of wealth flow, for instance, perceives children's contribution to labour within agricultural households as economically rational responses to familial wealth flow (Caldwell 1976). If the negative impacts of climate change on agricultural production reallocate labour towards agriculture and agricultural income increase due to scarcity of agricultural products, the higher return to working agriculture would result in higher fertility (Casey et al. 2019). Whilst the theoretical model calls for the consideration of the role of environmental change on demographic behaviour, empirical research on this issue is scarce (Hunter and Menken 2015). To date, there is no comprehensive and quantitative assessment of the impact of climate change on future population (Cohen 2010).

Indeed, what is missing from the empirical literature is the understanding of the direction, the mechanisms and the extent to which climate change affects and will affect demographic outcomes. The scarcity of scientific studies on the current and future impacts of climate change on population dynamics impedes the advancement of knowledge on future population trajectories and consequently hampers policy efforts to anticipate demographic challenges under future climate change. Future demographic research which addresses the following questions is fundamental in understanding the climate feedback on population allowing for more accurate inclusion of demographic variables representing the human systems in the integrated assessment models (IAM) of the IPCC (Jiang and Hardee 2011).

1) In what direction and to what extent does climate change influences fertility, mortality and migration – the three demographic components underlying population change?
2) How do the impacts of climate change on demographic outcomes vary by population subgroups?
3) What are the mechanisms through which climate change influences fertility, mortality and migration?
4) How does climate change affects future population size, composition and structure based on its effects on fertility, mortality and migration?

Whilst some research progress has been made regarding the climate impact on health and wellbeing, there is no scientific consensus regarding the direction and the extent to which climate change will influence population dynamics. The number of studies looking at the impact of climatic factors on fertility is limited and the focus is mainly on high income countries (Grace 2017). In particular, it is not clear whether climate change will lead to a fertility increase due to child mortality replacement (Kraehnert et al. 2019) or a fertility decline due to concerns about climate change (Arnocky et al. 2012; De Rose and Testa 2015). Regarding mortality, there is quite a number of studies on the effects of extreme events, particularly extreme temperatures or heat waves (Achebak et al. 2019; Martens 1998; Son et al. 2019) and exposure to hydro-meteorological hazards (Doocy, Daniels, et al. 2013; Doocy, Dick, et al. 2013) on mortality. However, these studies typically focus on one climatic factor or one specific type of natural hazard. As discussed above, a specific group of population is not vulnerable to all types of climatic hazards since it depends on their degree of susceptibility, risk exposure and adaptive capacity. There is thus a need for a synthesis of the evidence on climate-related mortality considering differential vulnerability.

With respect to the climate impact on migration, the past decade has seen a remarkable increase in the number of empirical studies focusing on climate or environmental driver of migration (Hoffmann et al. 2020). However, there is little empirical consensus concerning the direction and the extent to which these factors influence migration (Borderon et al. 2019; Piguet et al. 2018). Environmental change has been found to contribute to increase human migration in some studies but to constrain migration in others (Berlemann and Steinhardt 2017; Black et al. 2011; Borderon et al. 2019; Cattaneo et al. 2019; Hunter et al. 2015; Piguet 2010). There is also a lack of unified understanding about demographic profiles of migrants, i.e. who moves and who stays in the context of climate change. Without evidence-based knowledge and consensus in scientific findings, it is not possible to anticipate the direction and the magnitude of the effects of climate change on fertility, mortality and migration.

The lack of comprehensive scientific knowledge about how climate change is likely to impact future fertility, mortality and migration trends in different parts of the world makes it difficult to realistically forecast future population size and structures. The knowledge on future population trends and their vulnerability is crucial for the understanding of the socioeconomic costs of climate change and consequently for policy design. None of the two most widely used sets of global population projections factors in the possible consequences and associated discontinuities that may result from climate change. The United Nations (UN) population projections are based on an extrapolative hierarchical Bayesian model which essentially assumes that countries less advanced in the process of demographic transitions will follow the path of the more advanced countries (United Nations 2019). This, by definition, does not include possible new discontinuities such as those potentially caused by climate change. The expert argument-based global population projections by age, gender and level of educational attainment for all countries in the world up to 2100 produced by the Wittgenstein Centre for Demography and Global Human Capital and International Institute for Applied Systems Analysis (from now Wittgenstein Centre/IIASA) in 2014 (Lutz, Butz, et al. 2014) and 2018 (jointly with the Joint Research Centre (JRC) of the European Commission (Lutz et al. 2018) rely on judgements of over 550 population experts on a large set of predefined substantive arguments about the future drivers of fertility, mortality and migration trends. Although climate change was listed as a possible factor driving mortality and migration, the population experts surveyed in 2011/12 largely dismissed climate change as a relevant driver of population dynamics, presumably because not much research on these topics had been conducted by that time.

Indeed, the concern about the potential misrepresentation of future population is raised by Jones and O'Neill (2016) as a caveat in their global-scale, spatially explicit population projections. If people do move away from drought-ridden regions or low-elevation coastal areas facing rising sea-levels, for instance, not accounting for the climate driver of migration or urbanisation would mislead their spatial projections of future population size and distribution. If climate change does influence fertility, mortality and migration, the underlying assumptions for future population trajectories will have to account for the impacts of climate change on these demographic processes. Quantifying the effects of climate change on population dynamics would also lead to improvements in the predictive ability of forecast and projection exercises. Jones and O'Neill (2016) consequently highlight that considering the climate feedback on population dynamics should be given high priority in the future.

From a policy perspective, it is crucial to understand how exactly climate change will affect livelihood and wellbeing. The knowledge of the mechanisms through which climate change influences demographic and socioeconomic outcomes allows for designing an appropriate intervention on the right dimension. Agricultural production, food access, food prices and income losses are routinely referred to as key channels by which climate operates (Cattaneo and Peri 2016; Henry and Dos Santos 2013; Hunter et al. 2014). In the context of childhood undernutrition induced by floods and droughts exposure (Dimitrova 2020; Dimitrova and Muttarak 2020; Muttarak and Dimitrova 2019), for instance, knowing how these climatic hazards affect child health would allow for designing an appropriate intervention to alter the course of malnutrition. Despite the growing number of empirical studies on the impacts of climate change, there is a lack of understanding about the mechanisms by which climate affects demographic and socioeconomic outcomes (Carleton and Hsiang 2016; Gemenne et al. 2014; Mueller et al. 2020). Indeed, there is a call for empirical research which address these issues. This can help foster policy efforts to modify the mediating factors in order to minimise risks associated with climate change.

Demographers already have methodological tools to empirically assess, quantify and forecast the impact of climate change on population trends. The empirical analysis of fertility, mortality and migration can include changing climatic conditions in a local area as an exogenous determinant of the demographic processes. Population projections can explicitly include the climate feedback on key demographic behaviour when building up assumptions about future population trends. The analysis of the mechanisms through which climate change influences demographic and socioeconomic outcomes can employ structural equation modelling which allows for explicitly identifying the structural relationships between all the variables included in the model as well as distinguish between direct and indirect effects. Therefore, technically, available statistical techniques and models in demography and other social sciences can handle the new research directions proposed.

As for the data sources, some existing relevant data can readily be applied. For these exercises, the data that are either geo-referenced or uniquely identifiable by geographical location are required. Examples of global gridded climate and natural disaster data include: 1) climate data from the Climatic Research Unit at the University of East Anglia available from 1901; 2) drought events from the Global Drought Observatory available from 1951; 3) flood events from Dartmouth Flood Observatory available from 1985; 4) natural disaster events from Emergency Events Database (EM-DAT). Demographic data with information on geographical locations can be sourced from various surveys. These include: 1) Demographic and Health Survey (DHS), nationally-representative repeated cross-sectional surveys with detailed information on population characteristics and maternal and child health; 2) Integrated Public Use Microdata Series International (IPUMS-I) database comprises census microdata for 102 countries available from 1960. Socioeconomic data which are particularly relevant for the analysis of mechanisms include, for example: 1) Normalized Difference Vegetation Index (NDVI), remote sensing-based vegetation measures used as an indicator of agricultural yields (Boschetti et al. 2009); 2) food prices data from the Food Price Monitoring and Analysis database which collects monthly prices for several food commodities in selected markets in low- and middle-income countries; 3) conflict data from the Uppsala Conflict Data Program (UCDP) recording worldwide violent conflict from 1989.

Given data availability, population scientists have the right skills and expertise to embrace this new research agenda. Such efforts will provide foundations to our knowledge about future fertility, mortality and migration under climate change. With climate change consequences already being felt coupled with an increasing levels of public concern and civic activism (Capstick et al. 2015; O'Brien et al. 2018), assessing the climate consequences on population trends is highly timely. This does not only contribute to shed light on scientific uncertainty about the direction and the magnitude of the impact of climate change on population dynamics but also provides scientifically sound evidence for policies.

# Conclusion

The centrality of human population in the environmental and climate system and the urgency to tackle climate change mark the importance of demography at this present time of climate crisis. Significant advances have been made to understand the human impact on the environment and the global climate. Not only do the estimates and forecasts of population size and growth allow for projections of future emissions and the corresponding climate change, but the knowledge on population composition and distribution also contribute to a more precise assessment of the human impact. Similarly, estimation and projections of population heterogeneity allow for the identification of vulnerable subpopulations and assessment of future societies adaptive capacity. Population projections which are consistent with the global socioeconomic development narratives – the Shared Socioeconomic Pathways (SSPs) – provide understanding of how different development policies shape mitigation and adaptation challenges (Hunter and O'Neill 2014). Spatially-explicit population estimates and projections at a local or regional scale are also useful in identifying the population exposed to climate risks. Although research on environmental and climate change remains relatively marginal compared to other topics in population studies, demographers have already made significant contribution to the understanding of the interrelationship between humans and the environment and the climate.

Regrettably climate change is already happening and it is not possible to reverse many of its catastrophic effects. Whilst drastic emissions reductions and low-carbon transformations are urgently required to limit global warming to below 2°C by the end of the century, enhancing adaptive capacity is also key to mimimise the climate impact on human populations. Since global climate change is here to stay, there is no better time for demographers to actively engage in climate research. With readily available data and advanced methodological and computational tools, in the coming decades, improved population projections which account for both demographic and spatial heterogeneity will greatly contribute to locate *where* vulnerable subpopulations live and *who* they are. Likewise, the comprehensive and systematic assessment of the impacts of climate change on current and future population trends would help the scientific community in building more realistic scenarios about populations trends under the rapid pace of climate change and inform the international debate over the social costs of climate change. With these new research agendas, the next generation of demographic research on environmental and climate change would greatly contribute to better social and environmental policies design and policy planning in the longer time horizon.